\documentclass[12pt]{article}

\usepackage[style=authoryear, sorting=nyt, backend=biber, natbib=true, maxcitenames=2]{biblatex}
\usepackage{graphics}
\usepackage{epstopdf}

\usepackage{psfrag}
\usepackage{amsmath}
\usepackage{amsfonts}
\usepackage{float}
\usepackage{lineno}
\usepackage{rotating}
\usepackage{url}
\usepackage{xcolor}
\usepackage[section]{placeins}
\usepackage[a4paper, total={6.5in, 10in}]{geometry}

\usepackage[english]{babel}
\usepackage[utf8]{inputenc}
\newcommand{\benum}{\begin{enumerate}}
\newcommand{\eenum}{\end{enumerate}}
\newcommand{\bize}{\begin{itemize}}
\newcommand{\eize}{\end{itemize}}
\newcommand{\eeq}{\end{equation}}
\newcommand{\beqa}{\begin{eqnarray}}
\newcommand{\eeqa}{\end{eqnarray}}
\newcommand{\beqsub}{\begin{subequations}}
\newcommand{\eeqsub}{\end{subequations}}

\definecolor{red}{rgb}{1,0,0} 
\definecolor{blue}{rgb}{0,0,0.8} 
\definecolor{green}{rgb}{0,0.5,0}

\title{
Stochastic Resonance in Climate Reddening Increases the Risk of Cyclic Ecosystem Extinction via Phase-tipping
}

\author{Hassan Alkhayuon\footnote{University College Cork, School of Mathematical Sciences, Western Road, Cork, T12 XF62, Ireland}, Jessa Marley\footnote{CMPS Department (Mathematics), University of British Columbia Okanagan, Kelowna, BC, Canada}, \\ Sebastian Wieczorek$^*$, Rebecca C. Tyson$^{\textrm{\dag}}$}

\date{\today}

\begin{document}

\maketitle

\begin{abstract}

Human activity is leading to changes in the mean and variability of climatic parameters in most locations around the world.
The changing mean has received considerable attention from scientists and climate policy makers. However, recent work indicates that the changing variability, that is, the amplitude and the temporal autocorrelation of deviations from the mean, may have greater and more imminent impact on ecosystems.

In this paper, we demonstrate that changes in climate variability alone could drive cyclic predator-prey ecosystems to extinction via so-called {\it phase-tipping (P-tipping)}, a new type of instability
that occurs only from certain phases of the predator-prey cycle.
We construct a mathematical model of a variable climate and couple it to two self-oscillating paradigmatic predator-prey models.  
Most importantly, we combine realistic parameter values for the Canada lynx and snowshoe hare with actual climate data from the boreal forest. 
In this way,  we demonstrate that critically important species in the boreal forest have increased likelihood of P-tipping to extinction under predicted changes in climate variability, and are most vulnerable during stages of the cycle when the predator population is near its maximum.
Furthermore, our analysis reveals that {\it stochastic resonance} is the underlying mechanism for the increased likelihood of P-tipping to extinction.

\end{abstract}

\section{Introduction}
\label{sec:intro}

\indent\indent The climate experienced in any local area is rarely consistent from one year to the next.  A series of dry years, will eventually be followed by wet ones, cold winters will eventually be followed by warm ones.  The rate of switching between these climate regimes can be plotted to determine the expected number of years when conditions will remain either in a {\it low productivity state}, or a {\it high productivity state}.  
Excessively long series  of years under one type of climate, such as that experienced during the extended drought of the Great Depression in North America, can occur but are unusual under typical variability. 

Much attention has been given to the fact that anthropogenic climate change (henceforth, climate change) is resulting in a gradual warming of global mean temperatures.  Climate change, however, is also altering the climatic {\it variability},  that is, climate variations on the scale of years (also called environmental stochasticity). Existing work has shown that alterations in climatic variability can have much greater and more immediate impacts on ecosystems than changes in mean climatic conditions \citep{hastings:2021, petchey:1997, yang:2019}.

Short-term variations in climate metrics (e.g. precipitation, temperature) can be quantified by their standard deviation, a measure of the amplitude of deviations from the mean, and their temporal autocorrelation, a measure of the typical frequencies  present.
One of the major alterations possible under climate change is a shift in the typical frequencies of climate variability.
More specifically, if the variability in local climates is analysed to determine the dominant frequencies present, climate change in many geographic locations is expected to lead to a downshift of these frequencies, or an increase in the  temporal autocorrelation of the climatic variability. 
In other words, climatic variability 
has a particular ``colour" (a non-uniform frequency spectrum) in any given location on the globe, and these frequency spectra are changing shape, on average tending to more ``reddened" -  or more autocorrelated - climatic variability \citep{dicecco:2018, lenton:2017,boulton2015slowing}.  
The amplitude component of climatic variability is also expected to increase with global warming, leading to greater extremes in  climate conditions.  Since extreme climate conditions tend to be detrimental or, at the very least, challenging for ecosystems, these extremes correspond to a further decrease in productivity during low productivity years.  
While local conditions can certainly follow different patterns of change,  most geographic regions are expected to experience an increase in both the autocorrelation and the amplitude of climatic variability \citep{dicecco:2018}.  The combined effect of these two changes in variability is that poor conditions will last longer, and be more detrimental, than would be typical under normal variability.
This scenario is the focus of this paper.

The effect of changes in climatic variability on population dynamics has become a matter of deep concern for ecologists~\citep{vasseur:2004, vasseur:2007a, vasseur:2007b, petchey:1997, yang:2019, hastings:2021}.  It has been well-established that environmental stochasticity has a strong effect on population dynamics, and so it is critical that we understand the effect of changes to this stochasticity.  Whether these changes will be beneficial or detrimental to a given ecosystem is a complex issue \citep{yang:2019, petchey:1997}, often with no clear answer \citep{barraquand:2013}.  


Recent work has demonstrated that increased reddening of climatic variability significantly increases extinction risk in populations whose dynamics include two stable states that are stationary, one with high population numbers and one with very low population numbers~\citep{vanderbolt:2018}. 
Here we consider the much less-studied case of a cyclic system of interacting populations with two stable states: A self-oscillating (or cyclic) coexistence state and a stationary extinction state. Below we describe the importance of oscillatory states in real ecosystems, and the challenges these states present under climate reddening.

High amplitude multi-annual oscillations around cyclic coexistence states are ubiquitous in consumer-resource systems, including predator-prey populations \citep{boutin:1995, hanski:1995, kendall:1999}.  
Cyclic predator-prey systems are common in the Northern Hemisphere \citep{kendall:1998}, and it is generally accepted that their oscillations are inherent in the interaction dynamics of the system, i.e., they are not simply the result of some external environmental forcing \citep{turchin:2003}.  That is, while external environmental factors certainly influence predator-prey cycles, these cycles would continue in the absence of such forcing.  

Many climate systems also exhibit either periodic oscillations or typical frequencies \citep{bathiany:2018}, which can have strong effects on demographic rates (births, deaths, predation, etc) \citep{paniw:2021, dejager:2020, bastillerousseau:2018, northfield:2013, svensson:2006}. The full ecological-climate system is thus an oscillator with an external forcing that has characteristic oscillatory components.  One such component is the variability in climate metrics.  A classic example is the variability in the typical frequency and intensity of the El Ni\~{n}o Southern Oscillation (ENSO), which is associated with the frequency and intensity of wet, cold winters on the eastern coast of the Pacific \citep{latif:2009}. Meaningful changes in the amplitude and/or frequency of climate metrics are being observed around the globe and are accelerating as climate change proceeds \citep{lenton:2017, hodgkins:2014}. These changes could have a strong effect on environments in which cyclic predator-prey systems currently exist \citep{bathiany:2018}.  

One important example is the snowshoe hare and Canada lynx predator prey system in boreal North America: This system exhibits high amplitude multi-annual cycles that affect the entire boreal food web, and is also subject to particularly rapid climate change \citep{hodgkins:2014}.  The snowshoe hare is a keystone species in the boreal forest, and so its survival is of critical importance to the entire food web, and to the ecosystem services it provides to human populations.  At the moment, however, it is unknown how the hare-lynx system will respond to changes in the irregular oscillatory forcing of climatic variability. 


It is well-known that externally-driven nonlinear oscillators can exhibit extremely complex behaviour \citep{andreas:2013, coullet:1992}, and so it is possible that changes in the colour and frequency of climate variability could put the persistence and stability of oscillating populations at risk. 
Here we refer to the stable oscillatory state as the base state, and to the extinction state as the alternative stable state. We are interested in the possibility of critical transitions from the base state to extinction due to changes in the colour and amplitude of climatic variability. 
Recent analysis of two classic predator-prey models \citep{alkhayuon:2021} shows that oscillatory predator-prey trajectories are vulnerable to collapse under sudden shifts between high productivity and low productivity climatic conditions, even if the oscillatory behaviour is stable under both climates.  
This new mechanism of collapse, which occurs only during certain phases of the cycle, is called {\it phase-tipping} or {\it P-tipping}.  This tipping mechanism is in contrast to the more classical  bifurcation-induced tipping or {\it B-tipping} that occurs when the  base state disappears in a dangerous bifurcation (e.g. a fold bifurcation), and extinction becomes the only stable state.  

\begin{figure}[t]
  \centering
    \includegraphics[width=0.45\textwidth]{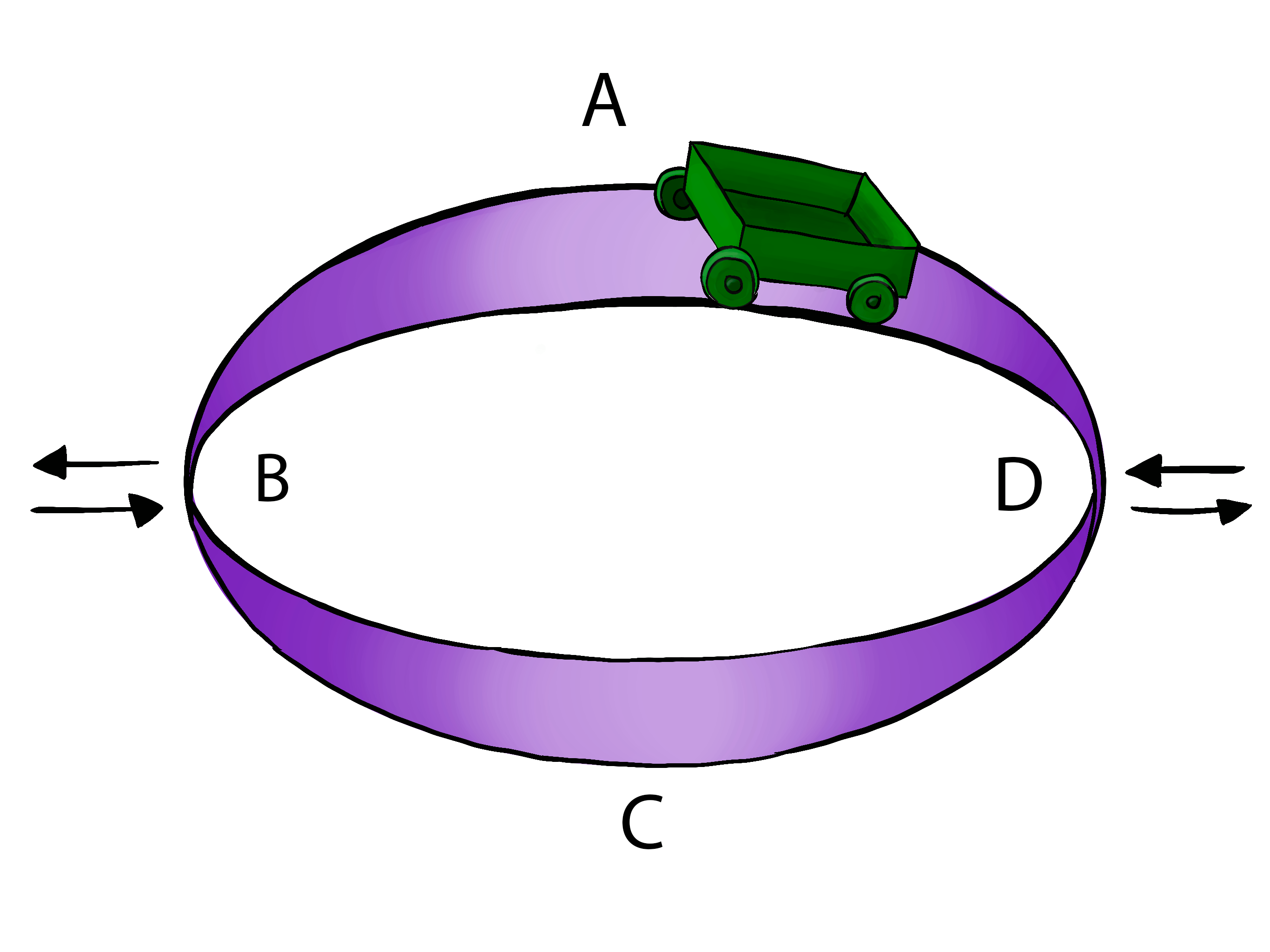}
\caption{Illustration of {\it phase-tipping} or {\it P-tipping}. 
The purple elliptical track represents the limit cycle, and the green roller-coaster car represents the predator-prey system. 
Climate variability results in movement of the entire limit cycle, as per the black arrows.  The car remains in contact with the track if the limit cycle shifts either right or left while the car is on side A or C of the track, or if the shift is left/right and the car is in position B/D.  
If the limit cycle shifts left/right while the car is in position D/B, the car will fall off the track.  So the car is only at risk of falling off the track if climate variability occurs in a particular direction and during particular phases of the rotation.  Hence, the name {\it phase-tipping}.  In this example the points A,B,C, and D represent four different phases of the cycle.} 
  \label{fig:p-tipping}
\end{figure}

Figure~\ref{fig:p-tipping} illustrates the basic mechanism of P-tipping, and the accompanying   {\it escape events} and {\it rescue events}.  
The purple track represents the limit cycle (the expected steady-state behaviour under constant climate).  It is a ribbon bent in a circle, which we are viewing from a slightly elevated angle, so that we can see the upper and lower halves of the circular track, but the sides appear skinny.  The green roller-coaster car represents the predator-prey system, which is oscillating as it follows the limit cycle around the circular track. 
Climatic variability results in a stochastic left-and-right movement of the entire track (limit cycle), as per the black arrows.[Note that changes in climatic variability alter the {\it shape} as well as the {\it position} of the limit cycle, and both play a role in determining those phases of the limit cycle that are vulnerable to P-tipping.  In Figure~\ref{fig:p-tipping}, however, we focus only on position changes, for the purposes of illustration].
If the climatic variability is small, and occurs while the car is 
on side  (phase) A or C, then, no matter which way the track moves, the car will remain on the track.  
That is, the predator-prey system will remain in the basin of attraction of the limit cycle.  
If the track shifts rightward while the car is at  phase D, the car also remains in contact with the track.  
If, on the other hand, the track shifts rightward while the car is at  phase B, the car will {\it escape}, i.e., disassociate from the track and fall outside the limit cycle basin of attraction. So the car is only at risk of falling off the track if climate variability occurs in a particular direction and during particular phases of the limit cycle.  The system may not have tipped, however, as it is possible, for at least a short interval, for climatic variability to shift the system far enough left to {\it rescue} the car and put it back in contact with the track.  If the escape event is not followed by a rescue event, however, then the car will never again be able to return to the track, and the system will have {\it tipped} to extinction. 

In comparison with the more extensively studied B-tipping, where the  base state disappears, extinction via P-tipping can occur in parameter ranges where the traditional understanding based on classical bifurcations would predict no vulnerability.  Thus, our work is demonstration in principle of a  ubiquitous but far less obvious mechanism for extinction that 
occurs even though the base state persists.  This mechanism
arises from intricate interactions
between the timescale of self-oscillatory predator-prey systems and the timescale of changes in the local climatic variability. 

Having identified this new tipping mechanism, we are led to ask: Will changes in contemporary climate variability interact with the oscillations of real predator-prey systems to trigger P-tipping to extinction?  If so, what is the nature of this interaction and the ensuing likelihood of P-tipping?   

We explore these questions using two paradigmatic predator-prey models with climatic forcing and
an Allee effect in the prey equation so that we can determine if extinction is deterministically possible. The predator-prey dynamic is parametrised using  the critically important Canada lynx and snowshoe hare system \citep{krebs:2001:book, peers:2020, krebs:2011}. Climate variability is represented as changes in the productivity rate of the prey, and is parametrised using climate data from the boreal forest in North America.

We show that P-tipping can indeed occur for realistic parameter values and changes in climate variability.  Furthermore, we show that tipping likelihood depends nonmonotonically on the colour of climatic variability, leading to the phenomenon of {\it stochastic resonance}  \citep{gammaitoni1998stochastic, berglund2022stochastic}. 
Our results suggest that the combined effect of anticipated alterations in the  climatic variability and stochastic resonance will result in an increased likelihood of P-tipping to extinction in the Canada lynx and snowshoe hare ecosystem in particular, and potentially other cyclic ecosystems in general.

\section{Models}
\label{sec:models}

Our study requires that we model two systems: The predator-prey system, and the climate system.  

For the predator-prey system we use two paradigmatic predator-prey models presented in \citet{alkhayuon:2021} and described in detail in Section~\ref{sec:pred-prey-models}.
To ensure relevance to real predator-prey systems, which is a key aspect of our work, we use realistic parameters of the extensively studied Canada lynx and snowshoe hare interaction.  
The snowshoe hare is a keystone species in the boreal forest, meaning that its survival is critical to sustaining the boreal forest ecosystem, and the Canada lynx is the hare's most important specialist predator.  Both species are famous for exhibiting high amplitude multi-annual oscillations \citep{krebs:2011, turchin:2001:book, krebs:2001:book}. 

For our climate model, we need a framework that allows us to control the variability in the climate time series, and couple it to the predator-prey system. Each variability component, that is, the amplitude and autocorrelation, responds to global warming in different ways.  We therefore model each component with a separate random process; see Section~\ref{sec:climate} for details.  Coupling to the predator-prey system is achieved through the prey productivity rate, which we consider to  be a function of randomly changing climatic conditions.
We use data from four locations in the boreal and deciduous-boreal forest of North America to determine the appropriate value of the autocorrelation parameter in our climate model; see Section~\ref{sec:transform-data}.

\subsection{Predator-prey models}
\label{sec:pred-prey-models}

We use the Rosenzweig-MacArthur and May (or Leslie-Gower-May) predator-prey models, both including an Allee effect, as formulated in \citet{alkhayuon:2021}.  We refer to these models the Rosenzweig-MacArthur-Allee (RMA) and May-Allee (MayA) models.  The variables $N$ and $P$ represent prey (snowshoe hare) and predator (Canada lynx) respectively.  We write the RMA model as 
\beqsub
\beqa
\dot{N} & = & r(t) N \left( 1 - \frac{c}{r(t)} N \right) \left( \frac{N-\mu}{\nu+N} \right) - \frac{\alpha NP}{\beta+N}, 
\label{eq:RM-prey}
\\
\dot{P} & = & \chi \frac{\alpha NP}{\beta+N} - \delta P,
\eeqa
\label{eq:RM}
\eeqsub
and the MayA model as 
\beqsub
\beqa
\dot{N} & = & r(t) N \left( 1 - \frac{c}{r(t)} N \right) \left( \frac{N-\mu}{\nu+N} \right) - \frac{\alpha NP}{\beta+N}, \\
\dot{P} & = & sP \left( 1 - \frac{qP}{N+\epsilon} \right).
\eeqa
\label{eq:May}
\eeqsub
The time-varying productivity rate [In the absence of the Allee effect, $r(t)$ is the prey intrinsic growth rate, i.e., the prey growth rate at low prey density.  With the Allee effect included in~\eqref{eq:RM-prey} however, $r(t)$ is no longer the intrinsic growth rate.  Nonetheless, this name continues to be used in the scientific literature, and there is no established name for the new role of $r(t)$ in models with an Allee effect.  As a compromise, we have chosen here to call $r(t)$ the prey productivity rate], $r(t)$, is directly proportional to the intrinsic growth rate and thus the carrying capacity.  Increasing $r(t)$ indicates a change in climate leading to higher productivity.  In the model, these conditions constitute an increased ``intrinsic growth rate" and environmental carrying capacity.  

The Allee term, $(N-\mu)/(\nu+N)$, means that the predator-prey system is vulnerable to extinction if hare densities become sufficiently low, but approaches 1 as hare densities increase.  The simpler and more common Allee term $(N-\mu)$ (no denominator) can create unrealistically large growth rates $r(t)(N-\mu)$ as hare density increases.  The form of the Allee term is chosen following \citet{drake:2006} and \citet{boukal:2002}.  While a demographic Allee effect has not been identified in snowshoe hare populations specifically, it has been identified in experimental predator-prey systems \citep{elliott:2017, kaul:2016, kramer:2010, gregory:2010} and in field data for some mammalian systems \citep{mclellan:2010}, and is widely viewed as a ubiquitous and important property of ecological systems across taxa \citep{kramer:2009}.

The parameter $\epsilon$ in the May model is introduced so that extinction ($N=0$) is nonsingular.  The implication is that the predator has alternate prey that supports a very low density predator population, but this density is so low that the predator can be considered extinct.  
The remaining parameters are briefly described in Table~\ref{tbl:params}, which also lists the values used.

 Both of these models exhibit cyclic solutions
, with 
trajectories rotating counter-clockwise  in the $(N,P)$-plane.  The corresponding prey and predator time series are periodic patterns of subsequent low, increasing, peak, and decreasing population densities. 
Each part of the cyclic solution is a  different phase of the oscillation, which can be precisely parameterised mathematically for
example, using the simple approach in \citep{alkhayuon:2021} or the concept of isochrons in \citep{guckenheimer1975isochrons}.

\subsection{Climate variability model}
\label{sec:climate}

 Our climate variability model is motivated by two key observations.  First, snowshoe hare demographic rates are affected by climatic conditions.  Second, variability in these demographic rates can remain close to zero for a few years, then can increase abruptly.  We discuss these two observations in more detail below.

Climate or weather has been shown to be linked to changes in demographic rates across many vertebrate species~\citep{wan:2022}, and researchers have investigated the effect of both global and local climate metrics on snowshoe hare populations.  Global metrics studied include the North Atlantic Oscillation (NAO), El-Ni\~{n}o Southern Oscillation Index (SOI), and the Northern Hemispheric Temperature (NHT) \citep{yan:2013, zhang:2007}. Local climate is measured with metrics such as the maximum temperature, days with snow cover, and snow depth \citep{peers:2020, burt:2017, meslow:1971, meslow:1968}.  
Correlations and even strong relationships can be found between hare demographics and a number of these climate metrics, and various mechanisms have been proposed, particularly for the local climate metrics. 
The system is complex, as most of these climatic factors are interrelated and can act on the predator-prey dynamics through multiple mechanisms \citep{lavergne:2021}.
Nonetheless, it is clear that several climate factors do have an effect on hare demographic parameters \citep{peers:2020, burt:2017, yan:2013, meslow:1971, meslow:1968},  particularly those related to population growth rate  (e.g.\  adult and juvenile survival, recruitment).  

Studies of the snowshoe hare show significant variations in demographic parameters, across large or small regions, with the mean value increasing or decreasing from one year to the next by an amount ranging from near zero to as much as $90\%$ of the previous year’s value. 

Among the parameter values that have been investigated, those with the strongest sensitivity include adult and juvenile survival, and measures related to fecundity (particularly number of litters and size of later litters) \citep{meslow:1968,meslow:1971,peers:2020,oli:2020}.
Finding what factors create these variations has been the subject of intense study. 
is that there is still considerable discussion over what exactly causes the hare-lynx cycle itself \citep{zhang:2007}, but the evidence indicates that density-dependent self-regulation \citep{zhang:2007}, predation \citep{oli:2020, yan:2013, turchin:2003}, cycle phase \citep{oli:2020},and climate \citep{burt:2017, peers:2020, yan:2013} are all important.  
 The relative importance of intrinsic {\it vs} extrinsic factors remains a matter of considerable debate, but it is nonetheless clear that the intrinsic factors still leave significant components of the variability unexplained, and that environmental conditions related to climate definitely play a role.
Thus, demographic rates can vary significantly from one year to the next  \citep{meslow:1968, meslow:1971} or from one cycle to the next \citep{oli:2020}, and it is entirely possible that climatic factors are important in these variations.

Our focus here is on the effect of changes in demographic rates due to changes in the {\it climate  variability} rather than changes in the mean. For simplicity, we focus on the prey productivity rate, $r(t)$, which can be directly linked to recruitment and juvenile survival, both of which are strongly affected by climatic conditions
\citep{meslow:1968,meslow:1971,yan:2013, peers:2020}.  To model climate-induced changes in demographic rates, several different approaches are possible.  Following \citet{wilmers:2007}, we adopt an approach that gives us access to both the amplitude and temporal autocorrelation of climate variability, and keeps the mean constant. Climatic conditions are divided into high productivity (type-H) and low productivity (type-L) years.  The degree of autocorrelation is given by the length of each period of type-L or type-H years. Increased autocorrelation thus leads to longer periods of a given type of climate (type-L or type-H), on average, before there is a switch to the opposite type. The climatic conditions are reflected in the value of $r(t)$.

For the amplitude component, we generalise the approach of \citet{wilmers:2007}, as we allow, in each scenario, a range of values for $r(t)$  selected from a closed interval $[r_{min}, r_{max}]$.
Sometimes, we use the {\it average} $r_{mid}$ and the {\it amplitude} $\Delta r$ instead, which are given by
\begin{equation} \label{eq:deltar}
    r_{mid} =\frac{r_{min} + r_{max}}{2}\quad\mbox{and}\quad \Delta r = r_{max} - r_{min}.
\end{equation}

Mathematically,  we use two random processes to model the variability. First, we assume that the amplitude of $r(t)$ is a random variable with a continuous uniform probability distribution on  the closed interval $[r_{\rm min}, r_{\rm max}]$. Second, the autocorrelation dictates the number of consecutive years $\ell$ during which the amplitude of $r(t)$ remains constant, and we assume that the length of this interval is a random variable with a 
discrete probability distribution known as the geometric distribution. In the statistical literature, the geometric distribution gives the number of failures in a sequence of independent Bernoulli trails until the first success occurs, where $\rho$ is the probability of success in each trial~\citep{Devroye2006}.
\begin{align}
g(\ell) = \mbox{Pr} (x = \ell) = (1-\rho)^\ell\, \rho,
\label{eq:geometric}
\end{align}
where $\ell\in\mathbb{Z}_+$ is a positive integer and  $\rho\in(0,1)$.
Together, these two processes give us an $r(t)$ that can be viewed as bounded autocorrelated noise.  

\begin{figure}[H]
  \centering
    \includegraphics[width=0.8\textwidth]{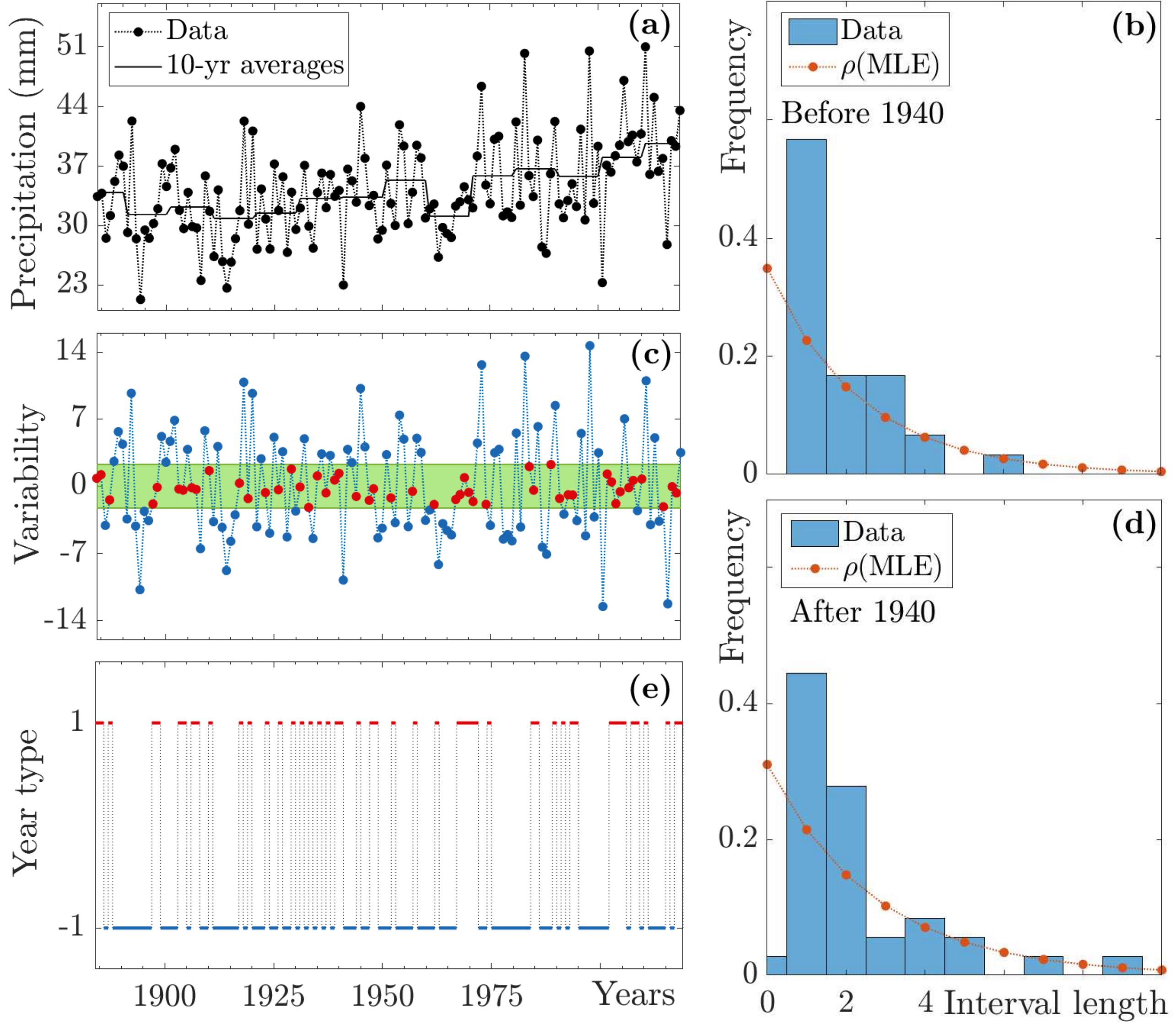}
    \caption{Left Column, Precipitation data from Burlington, VT (USA). Plot (a): Annual precipitation from 1884 to 2019 (black dots) with means calculated over 10-year intervals ( solid black).  Plot (c): Precipitation data with means removed, leaving the variability.  The data values within/outside 
    the green band correspond to high/low productivity years (red/blue dots).  Plot (e): Conversion of the data into binary form, 1 denoting high (red) and -1 denoting low (blue) productivity periods.  Right Column: Histograms of the length of high and low productivity periods, with the geometric distributions for the most likely value of $\rho$ computed using maximum likelihood.
    Plot (b): pre-1940 ($\rho=0.349$).  Plot (d): post-1940 ($\rho=0.310$).
    }
  \label{fig:BurlVTdata}
\end{figure}

The climate simulation algorithm proceeds as follows.  Suppose that the initial simulation time is $t=t_0$.  The algorithm first selects $\ell$ according to~\eqref{eq:geometric}, and then selects $r$ from a uniform distribution on $[r_{min},r_{max}]$.  Then the productivity parameter is set to $r(t)=r$ for $t_0 \leq t \leq t_0+\ell$, and the simulation is run (with constant parameter values) for $\ell$ years.  Then, new values of $\ell$ and $r$ are selected, and the simulation is run for $\ell$ years.  This process repeats until either the prey population goes extinct or the maximum simulation time is reached.

To determine realistic values for $\rho$, we examine climate data from the boreal and boreal-deciduous regions in North America where Canada lynx and snowshoe hare populations are found. 
Very few weather stations in this region have uninterrupted [Time series missing the occasional isolated data point could be made ``uninterrupted" by replacing the missing data point with the mean of the data points on either side] data for at least 100 years.  
We analysed precipitation data from Burlington, Vermont (USA), Fairbanks, Alaska (USA), and temperature data from Pelly Ranch, Northwest Territories (Canada), and Montreal, Quebec (Canada).  In order to determine the geometric switching rate, $\rho$, we first had to transform each time series by removing the mean, and classifying conditions as type-H and type-L.  Below, we describe this work.

\subsection{Determining the geometric switching rate}
\label{sec:transform-data}

All four data sets show a strong increasing trend over all or part of the time series.   
We compute the mean of the data over decades (adjacent 10-year windows), and remove these means from the data, leaving an adjusted time series containing just the variability. 
In order to estimate $\rho$, we then need to convert each adjusted time series into a binary time series where $+1$ corresponds to ``high productivity climate'' and $-1$ corresponds to ``low productivity climate''.  
What constitutes a ``low'' or ``high'' productivity climate is likely to vary from one organism to another, and from one location to another.  We define ``high" and ``low" productivity climates as corresponding to, respectively, data values inside and outside a ``normal variability" band around the mean.  
This definition is consistent with the notion that mean values are best, and when variability extends outside the ``normal variability" band, in either direction, productivity is decreased (see supplementary material for details). 

With this binary time series of type-H and type-L conditions, we can determine the length (in years) of each interval of constant conditions (type-H or type-L).  These measurements are then used to create a histogram showing the frequency with which each each interval length appears in the data.  Finally, we use Maximum Likelihood Estimation to fit~\eqref{eq:geometric} to this histogram and find the best-fitting estimate $\hat{\rho}$ of the probability ${\rho}$ (see supplementary material for details).

\begin{figure}[H]
  \centering
    \includegraphics[width=0.8\textwidth]{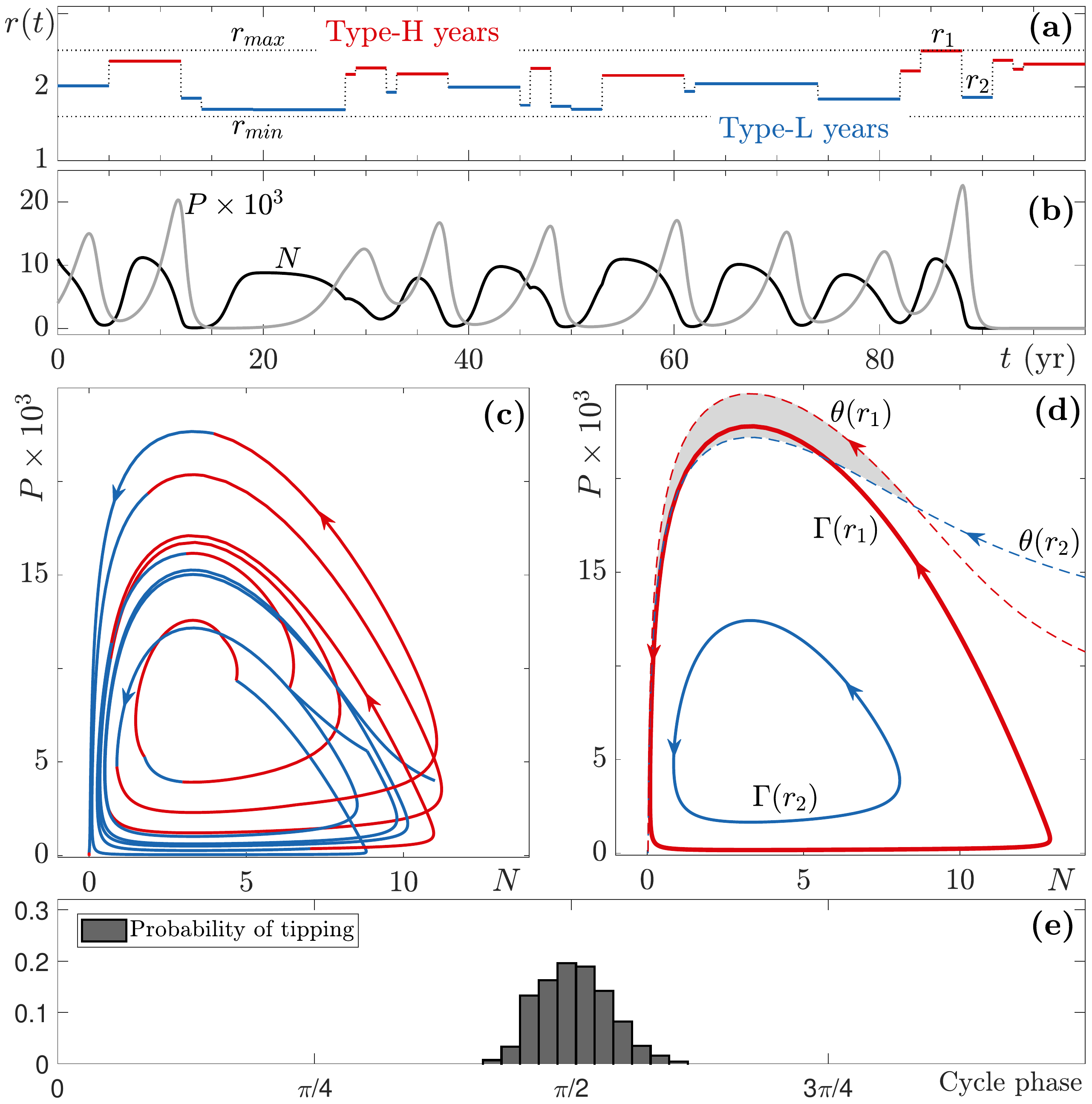}
\caption{
RMA model with switching climate. 
Plot (a): Climate as represented by $r(t)$ with $\rho=0.2$, $r_{min}=1.6$, and $r_{max}=2.5$. Type-H/type-L years shown in red/blue. Plot (b): Prey (black) and predator (grey) populations as functions of time.  Extinction occurs at approximately 90 years. Plot (c): Predator-prey phase plane trajectories for the time series in (b), with type-H/type-L segments shown in red/blue. Plot (d): Predator-prey phase plane under constant climate ($r(t)$ fixed).  The limit cycles $\Gamma(r_i)$ obtained under  $r=r_{1}= 2.49$ and  $r= r_{2}= 1.859$ are shown in red/blue ($ r_{1}$ and $ r_{2}$ are shown in (a)).  The corresponding basin boundaries $\theta(r_i)$ separating the limit cycle (below) and extinction state (above) basins of attraction are shown as dashed curves. Plot (e): Probability density of tipping phases computed using the Monte Carlo method with $10^3$ tipping experiments.
Other parameter values are given in Table~\ref{tbl:params}.
}
  \label{fig:RM_Tseries}
\end{figure}

Since we are interested in obtaining estimates for $\hat{\rho}$ before and during climate change, we seek to divide the data into these two respective periods.  
An increasing trend in the mean is evident in the latter part of the longer data sets we used, but a ``start" date for this increase is difficult to discern.  Globally, studies show clear evidence of anthropogenic climate change beginning to be apparent in local climate metrics at about 1940 \citep{hawkins:2020, hegerl:2019}. 
While this date can vary from one location to another, the 1940 transition date appears consistent with our data. 
We therefore choose 1940 as the year that defines the ``start" of anthropogenic climate change.  
To define the ``normal variability" band, we use the pre-1940 data, and set the band half-width (amplitude) as equal to $20\%$ of the maximum deviation from the mean (``before" climate change).  
To measure $\hat{\rho}$, we divide the data set into two parts, pre- and post-1940, in order to obtain values for $\hat{\rho}$ corresponding to the periods ``before" and ``during" climate change.

In Figure~\ref{fig:BurlVTdata} we show the results for Burlington, VT, which is the location with the longest climate time series.  The results for the other locations appear in the Supplementary Material.

\subsubsection{Example: Burlington, VT}
\label{sec:climate-data}

For Burlington, VT, annual precipitation data
for 1884-2019 is available from NOAA \citep{noaa:2020}.  The raw data, transformed
data, and switching interval histograms with geometric distributions
are shown in Figure~\ref{fig:BurlVTdata}.

We use Maximum Likelihood Estimation (MLE) to obtain the value of $\rho$ that best fits the data (see Supplement B for details). The maximum of the likelihood function applied to all of the data occurs at $\rho\approx0.32$.
The MLE values for $\rho$  corresponding to the other three locations (Fairbanks, Pelly Ranch, and Montreal) are similar, suggesting that $\rho$ between $0.3$ and $0.35$ is appropriate for climatic autocorrelation in the boreal and deciduous-boreal forest over the last one and a half centuries. Separating the pre-1940 and post-1940 data points for Burlington, we find that $\rho$ decreases from 0.35 to 0.31, consistent with the prediction of increasing autocorrelation projected under climate change.   In our work below, we find it helpful, in terms of developing our understanding of the system, to investigate the likelihood of P-tipping across the range of $\rho\in[0.1,0.5]$ around the realistic estimate of $\rho\approx0.32$.

\begin{figure}[H]
  \centering
    \includegraphics[width=0.8\textwidth]{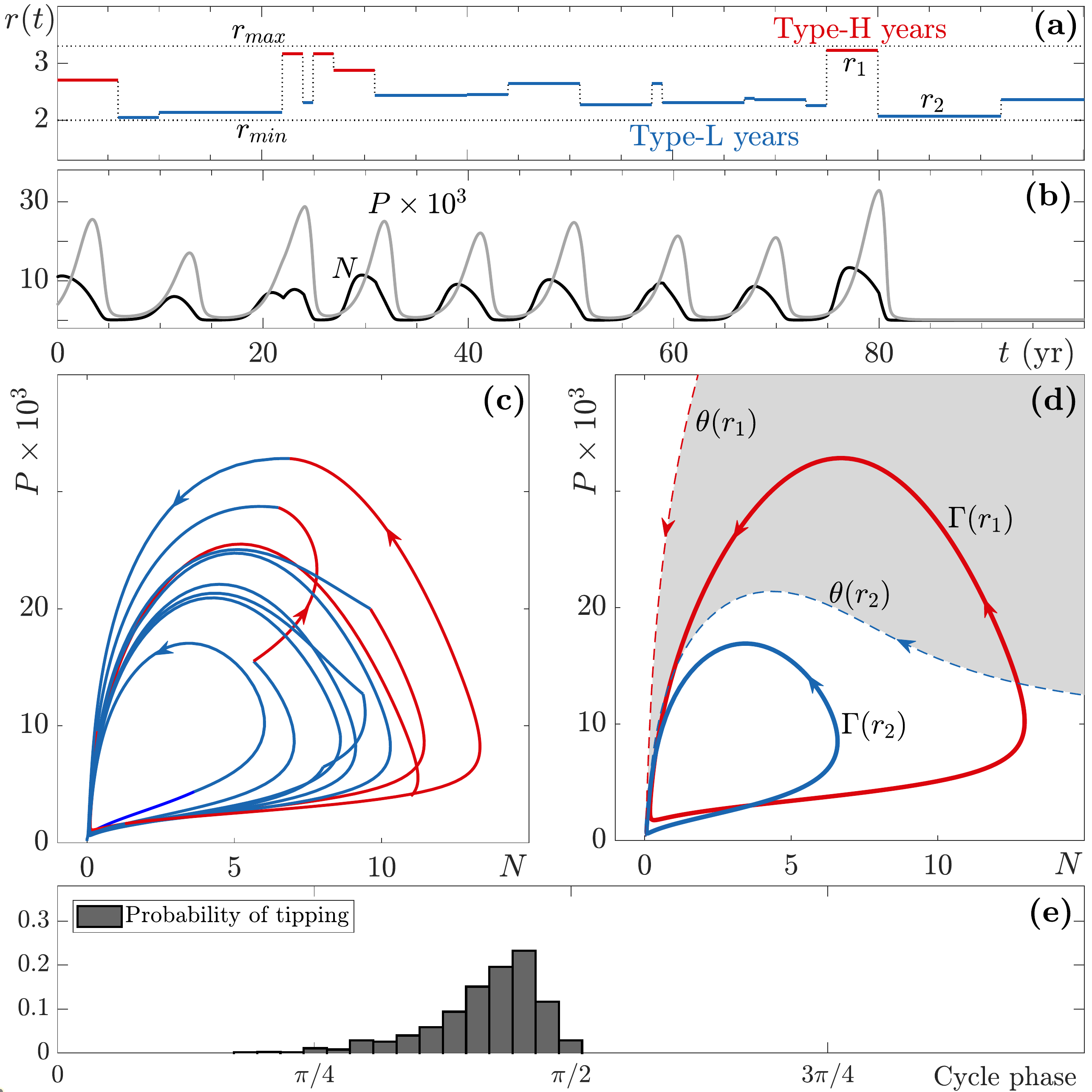}
\caption{MayA model with switching climate. 
Plot (a): Climate as represented by $r(t)$ with $\rho=0.2$, $r_{min}=2$, and $r_{max}=3.3$. 
Type-H/type-L years shown in red/blue. Plot (b): Prey (black) and predator (grey) populations as functions of time.  Extinction occurs at approximately 80 years. Plot (c): Predator-prey phase plane trajectories for the time series in (b), with type-H/type-L segments shown in red/blue. Plot (d): Predator-prey phase plane under constant climate ($r(t)$ fixed).  The limit cycles $\Gamma(r_i)$ obtained under $r= r_{1} = 3.22$ and $r= r_{2} = 2.068$ are shown in red/blue ($ r_{1}$ and $ r_{2}$ are shown in (a)).  The corresponding basin boundaries $\theta(r_i)$ separating the limit cycle (below) and extinction state (above) basins of attraction are shown as dashed curves. 
 Plot (e): Probability density of tipping phases computed using the Monte Carlo method with $10^3$ tipping experiments.
Other parameter values are given in Table~\ref{tbl:params}.
}
  \label{fig:May_Tseries}
\end{figure}

\section{Results}
\label{sec:results}

Below we describe the phenomenon of P-tipping to extinction and rescue events for cyclic predator-prey systems subject to varying climate in our model. We then move to the main topic of this paper: The
impact of predicted changes in climate variability, namely
increased amplitude  and 
temporal autocorrelation  of deviations from the mean \citep{ lenton:2017,dicecco:2018}, on the likelihood of P-tipping to extinction.

\subsection{P-tipping, escape, and rescue events}
\label{sec:results:phase}

The concept of P-tipping, illustrated in Figure~\ref{fig:p-tipping}, was introduced by \citep{alkhayuon:2021} as a new type of critical transition in nonlinear systems.
Here we explain this mechanism through  examination of the time series  of predator $P(t)$ and prey $N(t)$ populations, and the corresponding phase portraits in the predator-prey phase plane $(N,P)$. Sample time series are shown in Figures~\ref{fig:RM_Tseries} and~\ref{fig:May_Tseries}, for the RMA and MayA models, respectively.  All of the time series culminate in P-tipping.  Below, we give a detailed description of the subplots in these two figures, referring to both together.

As described in Section~\ref{sec:climate}, the productivity rate $r(t)$ varies stochastically between type-H (red) and type-L (blue) years, taking values within the range $[r_{min},r_{max}]$, according to changes in climatic conditions (subplots (a)).  Within this range of $r$, there is bistability between  extinction
and  the predator-prey  limit cycle.
The ensuing predator-prey oscillations
in subplots (b) have a timescale of approximately 10 years,
consistent with the Canada lynx and snowshoe hare population cycle in the boreal forest. 
In the phase plane,  the cyclic predator and prey populations trace out a counterclockwise trajectory composed of
red and blue segments, which correspond to type-H and type-L years of $r(t)$, respectively (subplots (c)).
To identify the key phenomena along this trajectory, we consider the limit cycle and its basin of attraction, set of all initial conditions $(P_0,N_0)$ that converge to the limit cycle for a fixed $r$,
for two different values of $r$ (subplots (d)).
In these subplots, we focus on two productivity values. These are the high productivity value $r_1$ and the low productivity value $r_2$, involved in the switch that triggers extinction. The triggering switch precedes extinction and occurs at approximately $t=88$ in Figure~\ref{fig:RM_Tseries}(b), and at approximately $t=80$ in Figure~\ref{fig:May_Tseries}(b). The limit cycle $\Gamma(r)$ (solid curve) is shown for $r=r_1$ (red) and $r=r_2$ (blue).
The basin of attraction of the red/blue limit cycle is the region below its red/blue basin boundary $\theta(r)$ (dashed curve). The region above $\theta(r)$ is the basin of attraction of the extinction state.
As the population trajectory in the phase plane follows the flow towards the limit cycle, there are parts or phases of the plane where the population is vulnerable to extinction if $r$ changes from $r_{1}$ to $ r_{2}$.
Note that, for every point in the basin of attraction of a limit cycle, one can define a phase of oscillations, for example, using the simple approach in~\citep{alkhayuon:2021} or the concept of isochrons in~\citep{guckenheimer1975isochrons}.
The vulnerable phases correspond to those portions of the basin of attraction of the high productivity limit cycle, $\Gamma( r_{1})$, that lie above the low productivity 
basin boundary $\theta( r_{2})$; see the shaded area in subplots (d)  and the probability density of tipping phases in subplots (e). Furthermore, we note that the vulnerable phases correspond broadly to those parts of the cycle where the predator population is near its maximum.  

\begin{figure}[H]
\centering
  \includegraphics[width=0.8\textwidth]{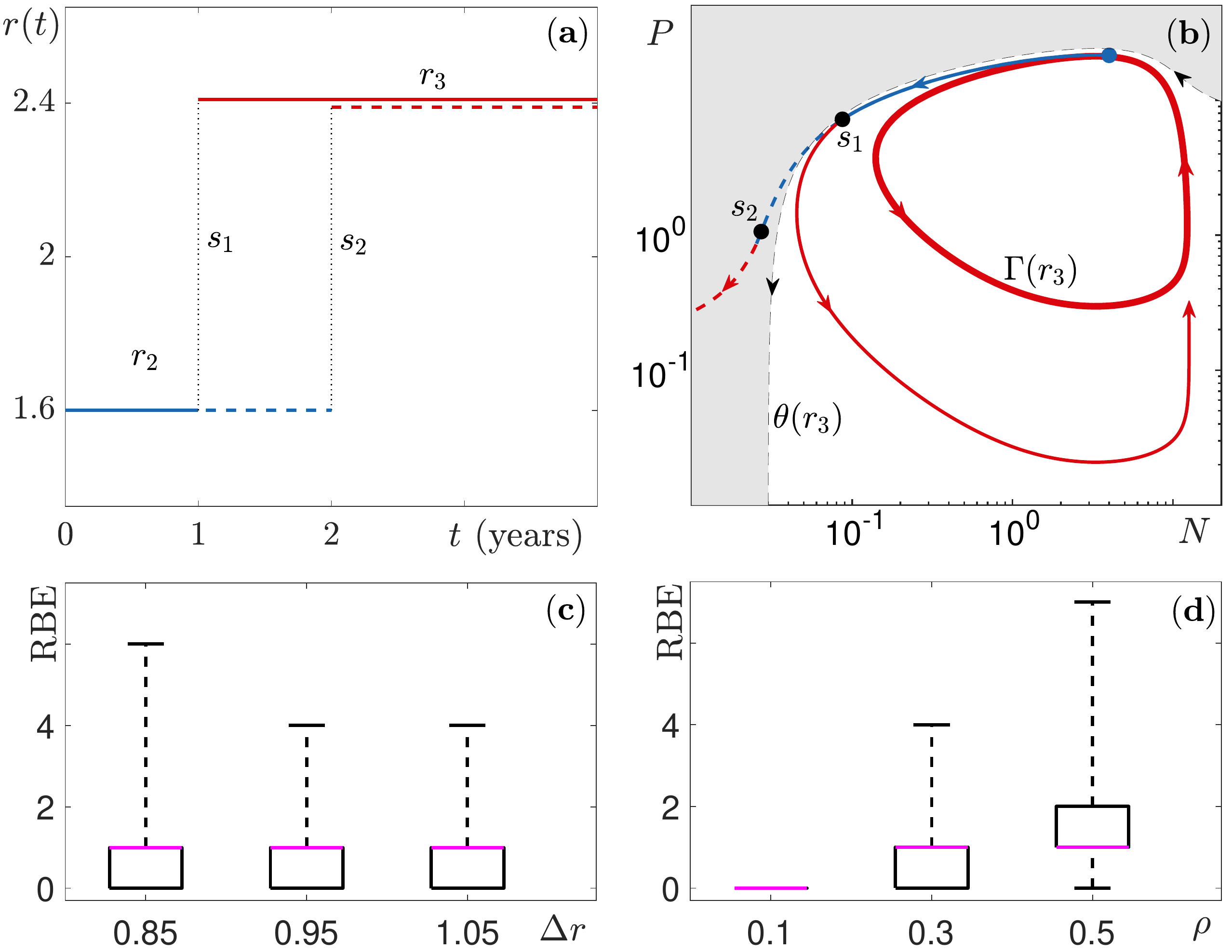}
  \caption{
    Rescue before extinction (RBE)  in the RMA model.  
    Plot~(a): Two time-varying functions $r(t)$, with the dashed red line jittered for readability; one $r(t)$ switches at  $t=1$ (solid), and the other at  $t=2$ (dashed).
    Plot~(b): An example of an escape event (blue dot) followed by either rescue or tipping to extinction, depending on the point where the switch from $r_2$ to $r_3$ occurs.  A switch at point $s_1$ ($t = 1$), results in rescue and thereafter the system (thin solid red curve) approaches  the limit cycle $\Gamma(r_3)$ (thick solid red loop).  A switch at point $s_2$ ($t=2$), results in tipping to extinction (dashed red curve).  The extinction point is not shown as the axes are logarithmic.
    The blue and red curves in (b) are obtained for $r=r_2=1.6$ and $r=r_3=2.4$, respectively. 
    The basin of attraction for the limit cycle $\Gamma(r_3)$ appears in white; the extinction state basin of attraction (for the same value of $r$) appears in gray.  
    Plots~(c,d): The number of RBE events for various values of $\Delta r$ and $\rho$. Plot~(c): $\rho = 0.3$.  Plot~(d): $\Delta r = 0.95$. All plots: $r_{mid} = 2.075$. Results are obtained using $10^4$ replicates at each value of $\rho$ and $\Delta r$. Other parameter values are given in Table~\ref{tbl:params}.
  }
  \label{fig:rescue_RM}
\end{figure}

Now, consulting either Figure~\ref{fig:RM_Tseries} or~\ref{fig:May_Tseries}, consider the following sequence of events.  Suppose that $r= r_{1}$ and the predator-prey system is on a counterclockwise trajectory approaching $\Gamma(r_{1})$.  
 When the trajectory enters the (shaded) vulnerable region,
the climate switches to $r_2$ conditions.  Consequently, the system suddenly finds itself in the basin of attraction of the extinction state.  This sudden switch from the basin of attraction of the limit cycle to the basin of attraction of extinction
is an {\it escape event}.  Once escaped, the system  moves towards extinction.  Nonetheless, this situation can be followed by two possible dynamical scenarios.
If the climate switches quickly enough
to some $r= r_3 > r_2$ such that the escaped system finds itself back in the  basin of attraction of the limit cycle, 
extinction  may be prevented. This recovery is called a {\it rescue event} illustrated in Figures~\ref{fig:rescue_RM}(a)~and~\ref{fig:rescue_May}(a).
On the other hand, if rescue does not occur, then extinction is inevitable and the system has {\it P-tipped}.  

To summarize: If there is a sudden and large enough drop in productivity while the system is in certain phases, and this drop lasts long enough,
the system is unable to adapt to the modified predator-prey limit-cycle and collapses to extinction. For more details about the phenomenon of P-tipping we refer to \citet{alkhayuon:2021}.

\begin{figure}[]
\centering
  \includegraphics[width=0.8\textwidth]{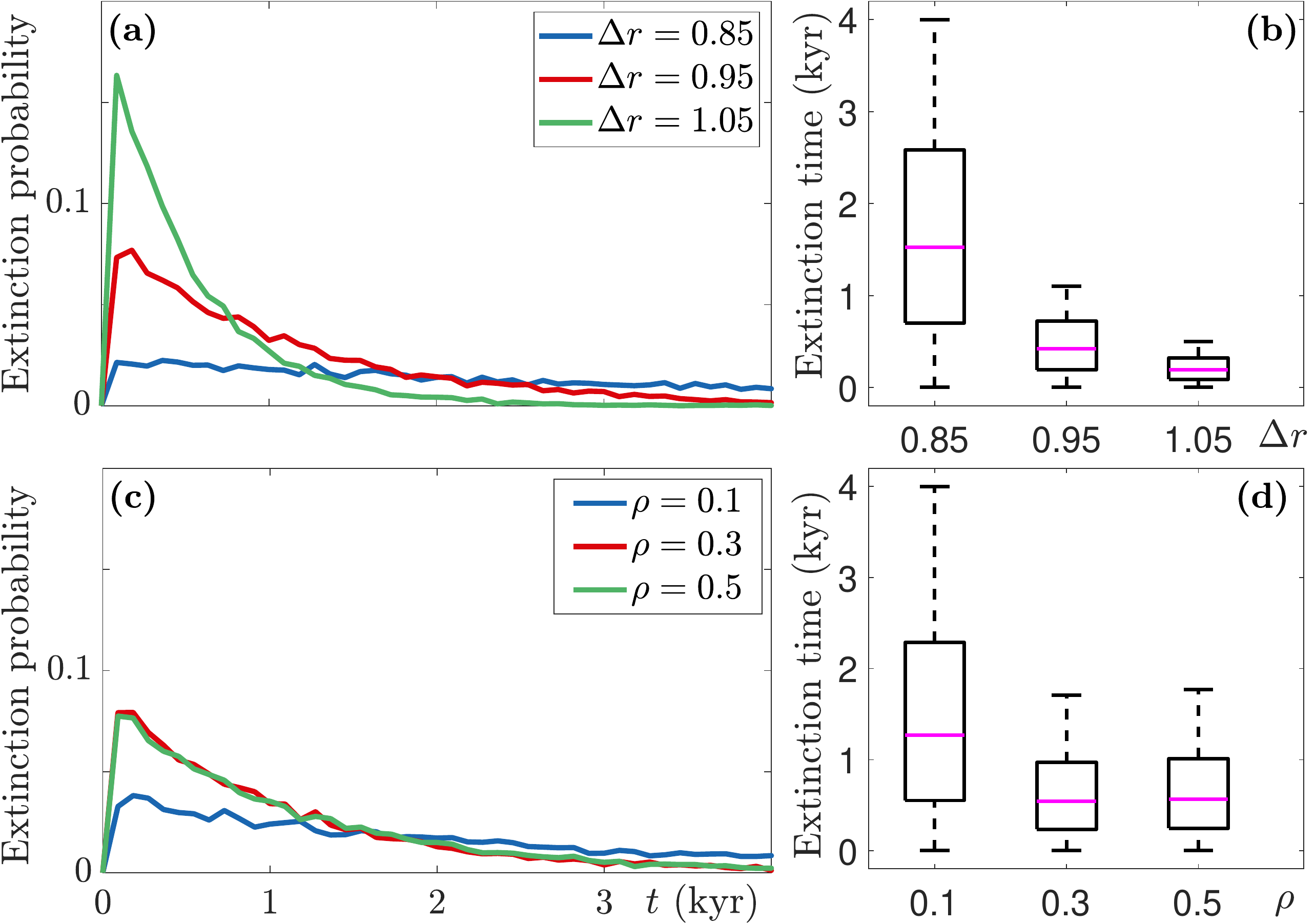}
  \caption{
  Distribution of the time to extinction (via P-tipping) for the RMA model.  Top Row: Distribution for different values of $\Delta r$.  Bottom Row: Distribution for different values of $\rho$. Left Column: Probability of extinction at time $t$.  Right Column: Box and whisker plots for the data in the left column.  For all plots $r_\textrm{mid} = 2.075$. See~\eqref{eq:deltar} for definition of $r_{\textrm{mid}}$.  Plots (a,b): $\rho = 0.3$.  Plots (c,d): $\Delta r = 0.95$. Results are obtained using $10^4$ replicates at each value of the parameter under investigation (i.e., $\Delta r$ in (a,b) and $\rho$ in (c,d)).
  Other parameter values are given in Table~\ref{tbl:params}
  }
  \label{fig:RM-Ptip-varypr}
\end{figure}

\subsection{ P-tipping likelihood under climate change}
\label{sec:results:warming}

In this section,  
we focus on the ecological implications of P-tipping on real predator-prey systems 
 under changes in climatic variability.
As the amplitude and autocorrelation of  environmental stochasticity
increase, we observe new interactions between the predator-prey and the climatic forcing 
In our model, larger amplitude corresponds to increasing $\Delta r$, and  increasing autocorrelation 
corresponds to decreasing $\rho$.

\begin{figure}[t]
\centering
  \includegraphics[width=0.8\textwidth]{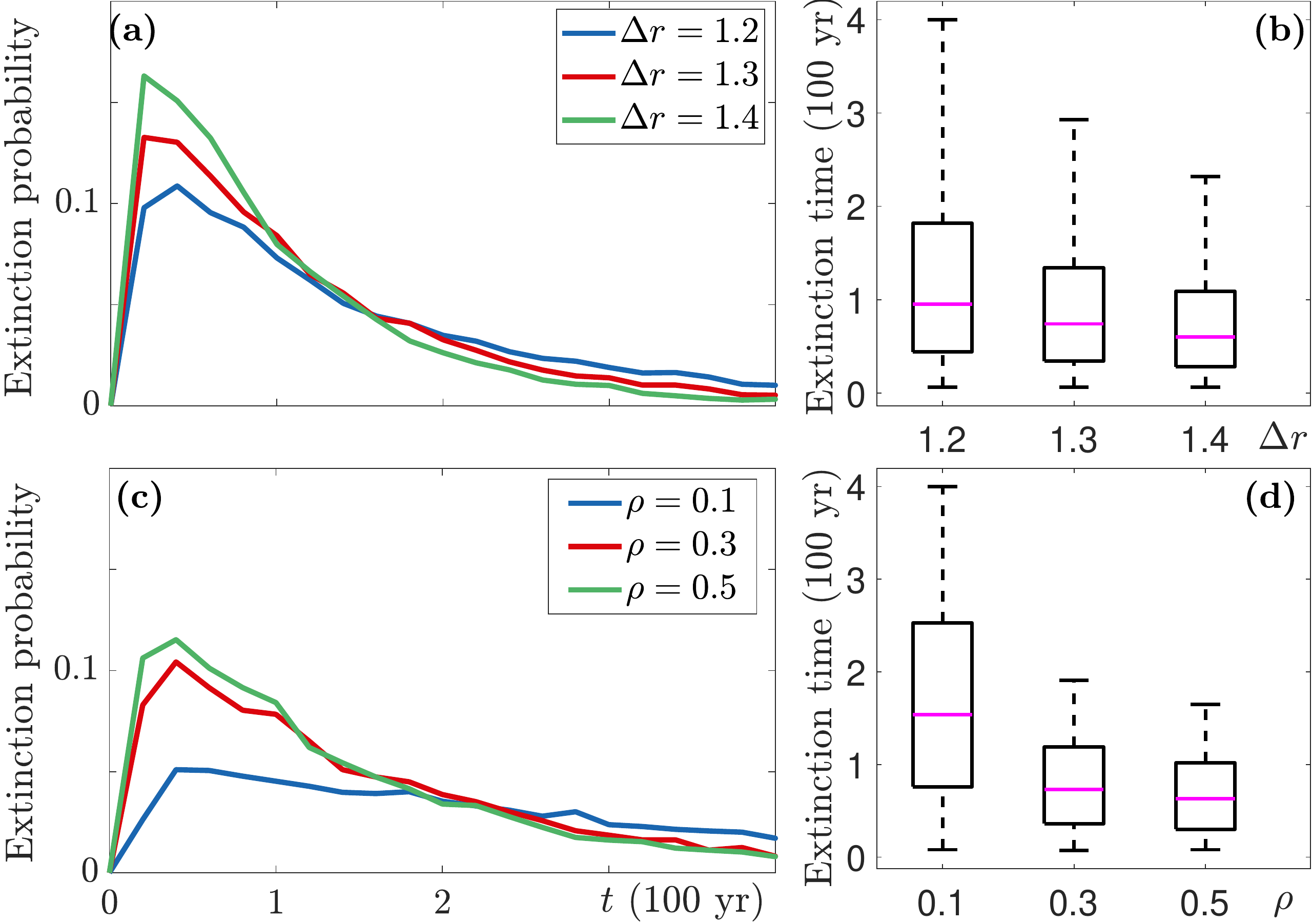}
  \caption{
  Distribution of the time to extinction (via P-tipping) for the MayA model.  Top Row: Distribution for different values of $\Delta r$.  Bottom Row: Distributions for different values of $\rho$. Left Column: Proobability of extinction at time $t$.  Right Column: Box and whisker plots for the data in the left column.  For all plots, $r_\textrm{mid} = 2.65$. See~\eqref{eq:deltar} for definition of $r_{\textrm{mid}}$.  Plots (a,b): $\rho = 0.3$.  Plots (c,d): $\Delta r = 1.3$. Results are obtained using $10^4$ replicates at each value of the parameter under investigation (i.e., $\Delta r$ in (a,b) and $\rho$ in (c,d).
  Other parameter values are given in Table~\ref{tbl:params}.
  }
  \label{fig:LGM-Ptip-varypr}
\end{figure}

In general we find that P-tipping is likely to happen sooner rather than later (Figures~\ref{fig:RM-Ptip-varypr} and~\ref{fig:LGM-Ptip-varypr}), except for very highly autocorrelated climates (small values of $\rho$).  In these cases the likelihood of tipping is relatively constant over all extinction times.  As noise amplitude increases, the likelihood of tipping sooner increases in both models, with the RMA model showing a marked increase in early P-tipping times.  The effect of increasing temporal autocorrelation is the opposite, leading to a decreased likelihood of tipping sooner.  So the predicted changes in climate variability, i.e., increased $\Delta r$ and decreased $\rho$, have opposing effects on expected P-tipping times, respectively causing a decrease and increase in expected extinction times.

We point out that the maximum simulation time in Figures~\ref{fig:RM-Ptip-varypr} and~\ref{fig:LGM-Ptip-varypr} is set to be $5\times10^3$ years. All simulations of the MayA model tip before reaching this time limit. On the other hand, up to $20\%$ of the RMA model simulations exceed the time limit without tipping. These simulations are considered safe from tipping and therefore excluded from the computation of the distribution of time to extinction.
That is, the maximum time to extinction shown reflects only those extinction events that occurred before the maximum simulation time. 

\subsubsection{Rescue events}
\label{sec:RBE}

\begin{figure}[H]
\centering
  \includegraphics[width=0.8\textwidth]{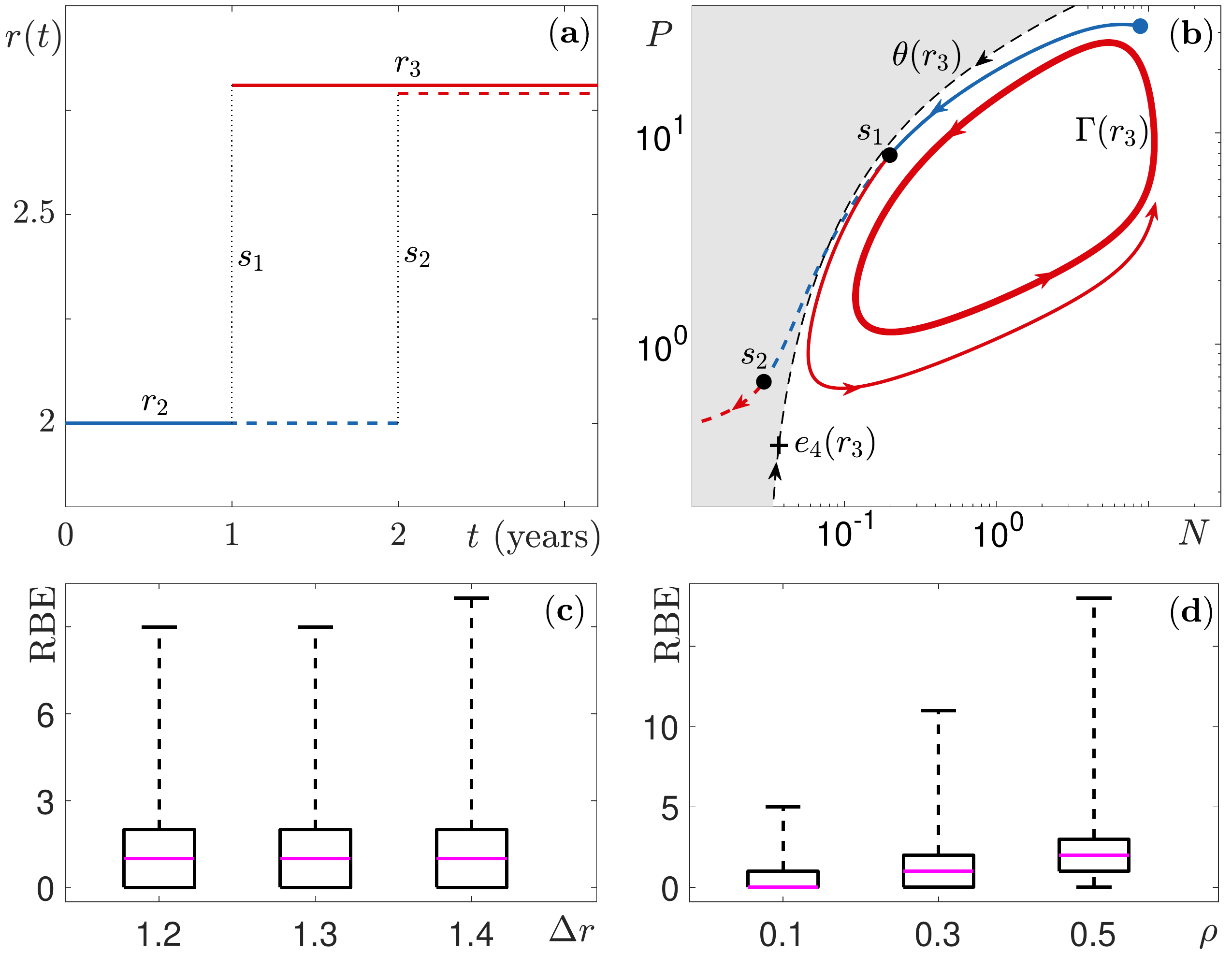}
  \caption{ Rescue before extinction (RBE) in the  MayA model.  
  Plot (a): Two time-varying functions $r(t)$, with the dashed red line jittered for readability; one $r(t)$ switches  at $t=1$ (solid), and the other  at $t=2$ (dashed).
  Plot~(b): An example of an escape event (blue dot) followed by either rescue or tipping to extinction, depending on the point where the switch from $r_2$ to $r_3$ occurs.  A switch at point $s_1$ ($t = 1$), results in rescue and thereafter the system (thin solid red curve) approaches  the limit cycle $\Gamma(r_3)$ (thick solid red loop).  A switch at point $s_2$ ($t=2$), results in tipping to extinction (dashed red curve).  The extinction point is not shown as the axes are logarithmic.
  The basin of attraction for the limit cycle $\Gamma(r_3)$ appears in white; the extinction state basin of attraction (for the same value of $r$) appears in gray.  
  The blue and red curves in (b) are obtained for $r=r_2=2$ and $r=r_3=2.8$, respectively.
  Plots (c,d): The number of RBE events for various values of $\Delta r$ and $\rho$. Plot (c): $\rho = 0.3$.  Plot (d): $\Delta r = 1.3$. All plots: $r_{mid} = 2.65$. Results are obtained using $10^4$ replicates at each value of $\rho$ and $\Delta r$. Other parameter values are given in Table~\ref{tbl:params}.
  }
  \label{fig:rescue_May}
\end{figure}

In Figures~\ref{fig:rescue_RM}~and~\ref{fig:rescue_May} we examine the number of rescue events that occur before the system finally P-tips to extinction.  
We call these events ``Rescue Before Extinction'' (RBE). 
First, we illustrate these rescue events in subplots (a) and (b). We are considering a situation where the system has just switched from $r=r_1$ to $r=r_2$. 
The initial state of the system is represented by a blue dot.
This dot lies inside the basin of attraction of the limit cycle $\Gamma(r_1)$ (not shown) but lies outside the basin of attraction of the limit cycle $\Gamma(r_2)$ (not shown), and so the system is heading toward extinction along the blue trajectory.
As discussed earlier (Section~\ref{sec:results:phase}), the system can still be rescued from extinction by another switch to $r=r_3$ sufficiently larger than $r_2$.
However, there is only a short time window during which this rescue event can occur. 
In both of these examples (Figures~\ref{fig:rescue_RM} and~\ref{fig:rescue_May}), rescue is possible if the switch from $r_2$ to $r_3$ takes place by time $t=1$ (subplot (a)), which corresponds to the switching point $s_1$ (subplot (b)).
In this case, the system follows the thin solid red trajectory approaching $\Gamma(r_3)$ (thick red loop, subplot (a)).  

Second, in subplots (c) and (d), we show how $\Delta r$ and $\rho$ affect the likelihood of RBE.
Our analysis shows that RBE is equally likely under changes in amplitude ($\Delta r$, subplot (c)), but more likely for less autocorrelated climatic  variability  ($\rho$ larger, subplot (d)). 
Our results provide some insight into the escape and rescue dynamics of the forced predator-prey system.  
Changes in the amplitude of climatic variability  have little effect on the likelihood of a rescue event occurring once an escape event has occurred. 
Increased autocorrelation of climatic variability, however, significantly reduces the likelihood that an escape event will be followed by a rescue event.

\subsubsection{Stochastic resonance}
\label{sec:SR}

In both the RMA and MayA models (see Figure~\ref{fig:resonance-heatmap}, second row) we find stochastic resonance \citep{gammaitoni1998stochastic}, i.e., a peak in  the probability of P-tipping to extinction at intermediate values of climate autocorrelation, (measured by the switching parameter $\rho$ (see~\eqref{eq:geometric})).  
Tipping probability is a consequence of the separate probabilities of escape and rescue events. The emergence of stochastic resonance indicates that
the likelihoods of escape and rescue events evolve differently under changes in climate colour.

\begin{figure}[t]
\centering
  \includegraphics[width=1\textwidth]{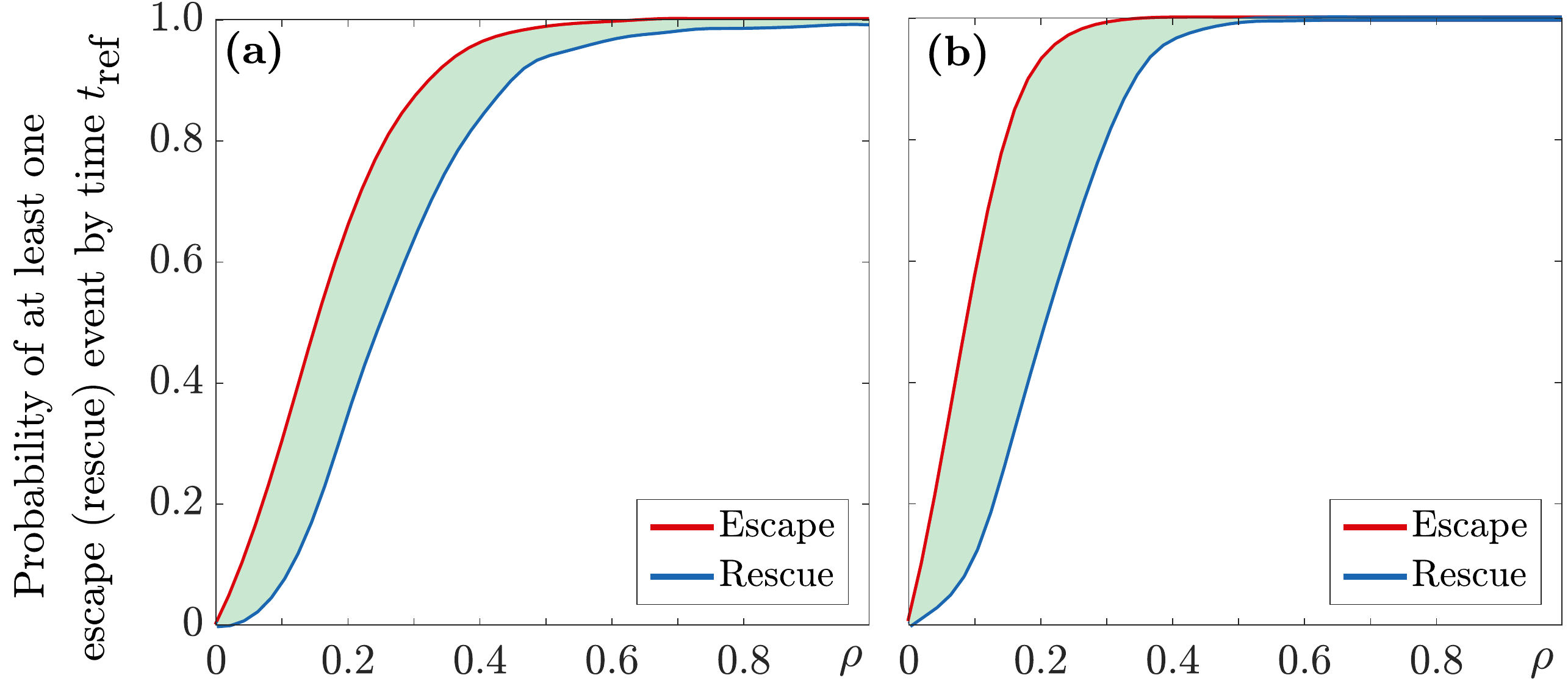}
  \caption{
  {Probability of at least one escape event (red) and least one rescue event (blue), taking place by time $t_\text{ref}$, as functions of $\rho$. 
  The green region indicates the difference between these probabilities. Plot (a): RMA model; $\Delta_r = 0.95$, $r_\text{mid} = 2.075$, and $t_\text{ref} = 550$. Plot (b): MayA model; $\Delta r = 1.3$, $r_{\text{mid}} = 2.65$ and $t_\textrm{ref} = 150$.
  Results are obtained using $10^3$ replicates for each value of $\rho$.
  Other parameter values are given in Table~\ref{tbl:params}.
  }}
  \label{fig:escape_and_rescue_prob}
\end{figure}

To understand the emergence of stochastic resonance, we describe the system behaviour as autocorrelation increases, beginning at very low levels  ($\rho\rightarrow 1$).
At low autocorrelation, the value of $r(t)$ changes frequently, and so escape events are highly likely. The rapid changes in $r(t)$, however, also make rescue events highly likely.  These two effects cancel each other out, resulting in a low likelihood of P-tipping. 
At intermediate levels of autocorrelation, however, the time scales of rescue and climate switching can interact.  
Climate switching generates both escape and rescue events, but there is an asymmetry in the way each event interacts with climate switching.  
Escape  events can occur at {\it any} switch from large enough to small enough values of $r(t)$, when the system is at the appropriate phase in the cycle,
but rescue events must occur within a short window of opportunity after the escape event. 
Thus, while both escape and rescue events become less likely as the autocorrelation increases, the likelihood of rescue  events is more strongly affected.  In particular, over any given time interval, the average length of type-L and type-H periods begins to exceed the time window over which rescue  events can occur.  The result is a significant decrease in the likelihood of rescue  events, and thus an increase in the likelihood of P-tipping.
Finally, extremely high autocorrelation means much fewer climatic switches between type-H and type-L years.  Consequently, escape  events are highly unlikely, and so is P-tipping, making the likelihood of rescue events essentially irrelevant. 
 We highlight this interaction in  Figure~\ref{fig:escape_and_rescue_prob}. 
We use Monte Carlo simulation to estimate the probabilities of at least one escape event (red) and one rescue event (blue) taking place by time $t_\textrm{ref}$ as a function the geometric parameter $\rho$. 
The difference between these two probabilities (green shaded region) shrinks in size as $\rho$ approaches $0$~or~$1$.

Our data suggest that the current value of $\rho$ is approximately $0.3$, which is worryingly close to the peak of tipping likelihood. 

\subsubsection{Net effect of climate change}
\label{sec:net-effect}

\begin{figure}[t]
\centering
  \includegraphics[width=0.8\textwidth]{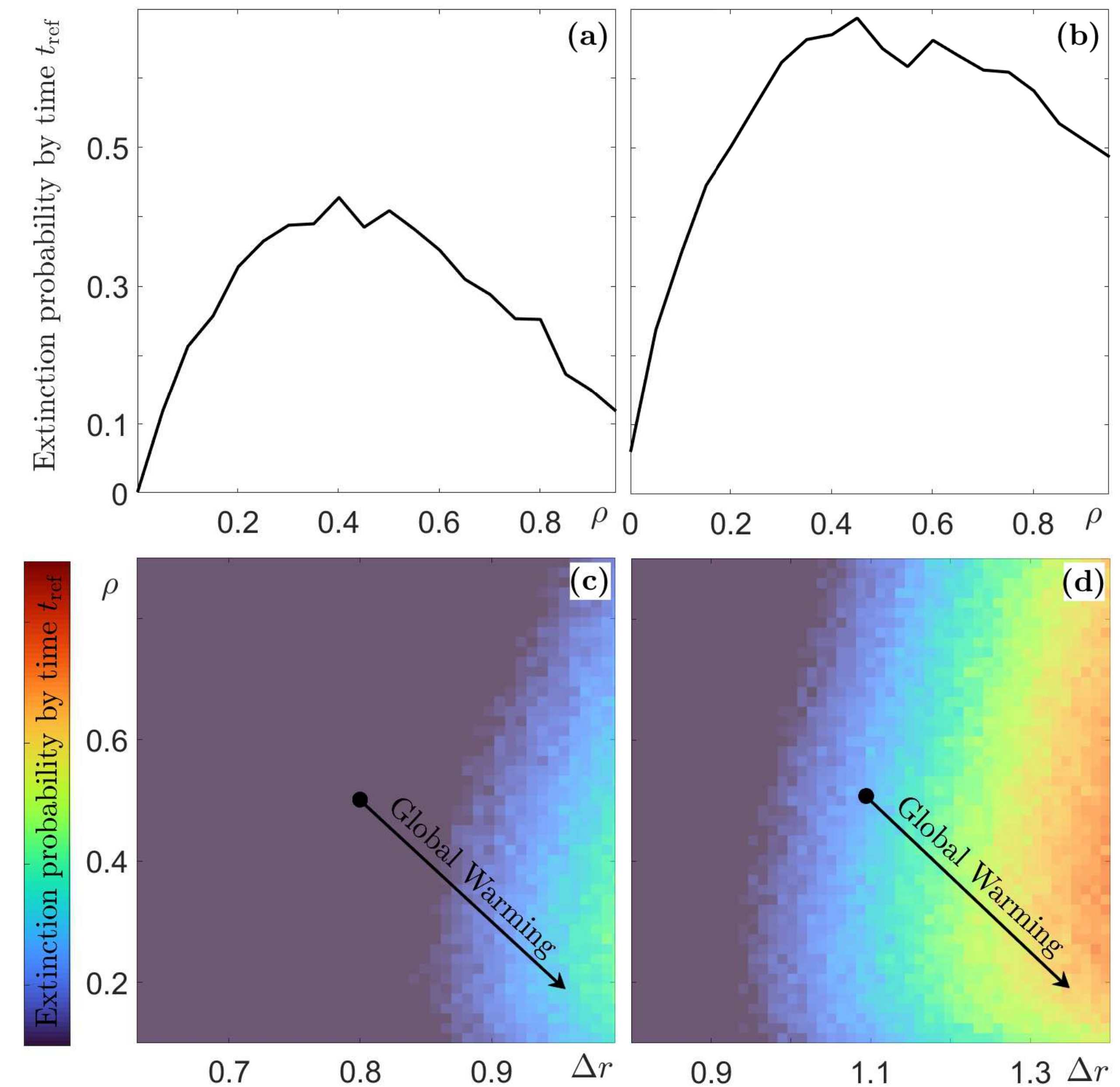}
  \caption{
  Probability of extinction taking place by a certain time $t_\textrm{ref}$. Left Column: RMA model.  Right Column: MayA model. Top Row: Extinction probability as a function of $\rho$ and $\Delta r$.  Plot (a): $r_{\text{mid}} = 2.075$ and $t_\textrm{ref} = 550$. Plot (b): $r_{\text{mid}} = 2.65$ and $t_\textrm{ref} = 150$.  
   Bottom Row: vertical cross-section of corresponding top plot showing stochastic resonance in the probability function on varying $\rho$ (stepsize in $\rho$ is 0.05) with fixed $\Delta r$. Plot (c): $\Delta r = 0.95$, $r_{\text{mid}} = 2.075$ and $t_\textrm{ref} = 550$. Plot (d): $\Delta r = 1.3$, $r_{\text{mid}} = 2.65$ and $t_\textrm{ref} = 150$.
  Results are obtained using $10^3$ and $10^4$ replicates, for (a)-(b) and (c)-(d) respectively, at each value of $\rho$ and $\Delta r$. 
  Other parameter values are given in Table~\ref{tbl:params}.
  }
  \label{fig:resonance-heatmap}
\end{figure}

We have shown that the likelihood of P-tipping generally increases with increasing amplitude of climatic variability, but that the effect of increased autocorrelation is non-monotonic. It is unclear, therefore, how these two effects will interact when they occur simultaneously, as is predicted under contemporary climate change.  We present these results in Figure~\ref{fig:resonance-heatmap}, top row.  
Under global warming, we expect the direction of change of $(\Delta r,\rho)$ to be as shown by the black arrow on Plots~(a) and (b).
While the actual trajectory will certainly be stochastic, and could exhibit slopes ranging from very steep to very shallow, it is nonetheless clear that the general trend will be a diagonal shift toward the bottom right corner.  The results suggest that the combination of increased amplitude and autocorrelation in climate variability will lead, generally, to an increase in the likelihood of P-tipping to extinction.

\section{Discussion}
\label{sec:discuss}

Species persistence in response to climate change has been studied extensively in the context of increasing local mean temperatures

\citep{vanselow:2019, culos:2014}, range shifts \citep{tejo:2016, zhou:2011}, and climate variability or frequency of extreme events \citep{huelber:2016, kaneryd:2012, hildebrand:2007, lynch:2014}.  More generally, the propensity of climate change to lead to critical transitions in populations (i.e., sudden dramatic changes in population size, generally toward much lower numbers) has been well-explained in models with bistability between two stationary base states \citep{dakos:2019}.  These two states could be, for example, a high population coexistence state and either a low population coexistence state or a state with extinction of one or both populations \citep{vanselow2022evolutionary, kaur2022critical, okeeffe:2019, vanselow:2019, osmond:2017}. However, little is known about the effects of climate change on systems whose base state is oscillatory \citep{sauve:2020, bathiany:2018}.

In this paper, we study the impact of {\it changing} climatic variability on the persistence of {\it cyclic} ecosystems, i.e., interacting populations whose base state is a {\it limit cycle}.  
In many parts of the globe, especially over the northern continents, global warming is generally leading to an increase in the autocorrelation and amplitude of climate metrics \citep{lenton:2017, dicecco:2018}.
In our model, changing climatic conditions are represented by changes in prey productivity rate, $r(t)$.
Increased autocorrelation and amplitude translate, respectively, to a gradual decrease in the climatic switching rate $\rho$, and a gradual increase in the switching amplitude  $\Delta r = r_{max} - r_{min}$.
Our starting values for $\rho$ are consistent with precipitation and temperature data from various locations in the boreal and deciduous-boreal forests of North America.  For $r_{min}$ and $r_{max}$, 
we use values that ensure continued existence of the predator-prey limit cycle throughout, meaning that there are no bifurcations that either destabilise or destroy the predator-prey cycle within the range $[r_{min},r_{max}]$.

The characteristics of our chosen $r(t)$ model that are key to our results are the presence of multiple years of fairly constant $r(t)$ followed by relatively large and rapid changes in $r(t)$.
This model was motivated by the data available for snowshoe hares, showing
both occasional large year-to-year variations in various demographic rates, as well as multiple years with similar demographic rates \citep{oli:2020, meslow:1971}. 
We used thresholds to translate the climate data into demographic rates, using the notion of ``good years" (close to the mean) and ``bad years" (far from the mean).  Climate time series have been well-studied and characterised, but there is little information about what time series are appropriate for demographic rates.  These parameters respond in a nonlinear fashion to climate inputs and are furthermore embedded in highly nonlinear food webs. Our approach is a simple first step in modelling this nonlinear process.
Furthermore, it is akin to
a series of press disturbances, a widely used approach where
each change in external conditions is modelled as a single switch
between two values of an input parameter \citep{schoenmakers2021resilience}.
One advantage of considering such a forcing profile is that the sudden shifts in $r(t)$ allow us to examine the changes in P-tipping likelihood relatively easily.

An alternative approach to modeling $r(t)$ would be to use an autoregressive model with degree $p$. 
Autoregressive modelling is a statistical technique often used to forecast a time series based on past behaviour \citep{box2015time}. It has also been employed to study sensitivity of plankton to temperature fluctuations \citep{beninca2011resonance}.
We have checked that with an autoregressive model for the demographic rate, P-tipping is much less likely, though it persists for sufficiently autocorrelated autoregressive models.
Autoregressive models have proven useful for understanding population variability \citep{lima:2002, ferreira:2016, salvidio:2008}, but rely on linearity assumptions that are not generally satisfied by ecological systems \citep{ives:2010}.
There has been no comparison of autoregressive models with other stochastic models for fit to demographic rates, at least in part because  
most ecological data sets do not allow for rigorous identification of a best climate model \citep{grosbois:2008}.  The existing data for snowshoe hares are insufficient to provide unequivocal support for an autoregressive model over our approach,
as a means of representing the response of demographic rates to environmental conditions. Furthermore, there is evidence that climate change may be altering the relative importance of density-dependent factors such as predation \citep{lavergne:2021}, putting more weight on the effect of local climate metrics \citep{lavergne:2021} rather than the time series history.

In order to 
ensure maximum relevance of our study to real ecosystems,
we frame our work around 
contemporary climate and a classic cyclic ecosystem: the oscillatory predator-prey system.
In particular, our climate parameter values are obtained from real boreal forest  climate data. For the ecosystem, we use parameter values consistent with the Canada lynx and snowshoe hare predator prey system \citep{strohm:2009}. 
In addition, we use two very different but widely accepted predator-prey models: the Rosenzweig-MacArthur (RMA) and the Leslie-Gower-May (MayA) models \citep{turchin:2003}.  Neither of these models is perfect for any particular real system, but together they provide an important baseline from which to investigate the dynamics of predator-prey systems \citep{turchin:2003}.  Indeed, both have led to insights into the Canada lynx and snowshoe hare ecosystem dynamics \citep{turchin:2003, strohm:2009, tyson:2009}. 
Our results show that the extinction risk via P-tipping, quantified in our previous paper under statistically constant climatic variability \citep{alkhayuon:2021}, is likely to be exacerbated under the predicted direction of changes in climate variability.  

For both the RMA and MayA models, $\Delta r$ has a positive relationship with the likelihood of P-tipping. 
Larger jumps in $r(t)$ correspond to larger changes in the size and shape of the limit cycle basin of attraction, and therefore increase the likelihood of escape events. That is, after a drop in $r(t)$ the limit cycle and its basin change shape in such a way that the system will be outside the basin of attraction of the changed limit cycle.
On the other hand, the mean time to extinction decreases significantly as $\Delta r$ increases, as shown in Figures~\ref{fig:RM-Ptip-varypr} and~\ref{fig:LGM-Ptip-varypr}.
Our work suggests that the maximum time to extinction also decreases with increasing $\Delta r$. 
[Note that, we limited our simulations to a maximum simulation time of $t=5000$ years.]

The relationship between likelihood of extinction and the climate switching rate, $\rho$, is more nuanced: Starting from high values of $\rho$, the likelihood of P-tipping initially increases until $\rho\approx0.5$, and then decreases, giving rise to {\it stochastic resonance}.  Our data suggests that the current value of $\rho$ is most likely in the range [0.1,0.5], which includes the stochastic resonance peak. %
We identify escape and rescue events in Section~\ref{sec:results:phase}, and show in Section~\ref{sec:SR} that
the stochastic resonance we observe is due to the differences in the way escape and rescue respond to changes in $\rho$. 
Recently, there has been much interest in the possibility of overshooting tipping-point boundaries for a short period of time without actually prompting dangerous transitions to alternative stable states \citep{ritchie2021overshooting}. Rescue events are a new example of this  phenomenon.

The fact that we observe P-tipping in two models with widely differing predator growth mechanisms provides strong indication that P-tipping is a phenomenon that exists across different predator-prey systems.  
The comparison also reveals that the likelihood of P-tipping can vary enormously: For the Canada lynx and snowshoe hare parameter values, subject to the same climate switching rate $\rho$ and climate amplitude $\Delta r$ (held within a range where no bifurcation exists), we find that P-tipping is as much as 10 times more likely in the MayA model than in the RMA model (compare Figures~\ref{fig:RM-Ptip-varypr} and~\ref{fig:LGM-Ptip-varypr}).  The increased complexity in more realistic models will certainly affect the likelihood of P-tipping, but results from studies of other tipping phenomena suggest that complexity may \citep{brook:2013} or may not \citep{klose:2020, kroenke:2020} be protective.

At this point, we are only able to point to the possible presence of P-tipping and stochastic resonance resulting from changes in climate variability, and so there is a need for future work in this area.  The two mechanisms that are key to the occurrence of P-tipping in our predator-prey system are the Allee effect and the type of stochasticity present in the prey demographic rates.  Both are likely to respond in different ways to warming-induced changes in climate variability, but little is known about these responses \citep{berec:2019}.  Other parameters may also be important, such as predator demographic rates or predation rates.  To simplify the model analysis, we considered only the case where the prey productivity rate $r(t)$ responds to changes in climatic conditions.  Further work is needed to determine the manner in which other parameters are altered by changes in environmental conditions and their effect on the likelihood of P-tipping.  In particular, the proportion of cycle time spent in the peak (the basin-unstable region) can vary considerably as model parameters are changed, and this is precisely the part of the cycle that is vulnerable to P-tipping.  To what extent we need to be concerned about P-tipping in real systems thus remains an open question.

 There is also a need for climate data sets that are longer and that originate from many different locations around the globe. 
The climate reddening currently predicted is largely at the planetary scale \citep{boulton:2015}; there has been less investigation of this phenomenon at local scales.  Recent work clearly shows that temporal autocorrelation is increasing in some regions while decreasing in others \citep{dicecco:2018, lenton:2017}.  
Consequently, the local character of climate variability, and the local ecological effects (i.e., the model parameters affected), are important factors to consider.  Here we modeled climatic conditions via changes in prey productivity rate, $r(t)$, but there are other parameters that could also be linked to climatic conditions \citep{marangon:2020, peers:2020}.
Furthermore, population dispersal could connect neighbouring regions that are suffering very different changes in climate variability \citep{petchey:1997}, leading to questions about the effect of spatial coupling on the likelihood of P-tipping.

While we find only a small likelihood of P-tipping in the models presented here, there are a number of factors that make this phenomenon concerning for management of real systems.
First, the 10-fold difference in tipping likelihood between these two models indicates that other models could have a much higher likelihood of P-tipping.
In addition, when coupled with other mechanisms such as tipping cascades, P-tipping could be an important factor increasing the overall risk of extinction.  Finally, no early warning signals have been identified for escape-type events, which occur far from a stable equilibrium point.
As a result, if a system is vulnerable to P-tipping, managers will have no way of knowing if the ecosystem is approaching an escape event.
Together, these observations highlight P-tipping as an important mechanism to investigate in studies of ecosystem resilience.

{
\section*{Data Accessibility}
The codes used to conduct simulations and  generate figures are available via the GitHub repository \citep{alkhayuon:2021repo}
}

\section*{Acknowledgements}  HA  and  SW  were  funded  by  Enterprise  Ireland  grant  no.  20190771.  RCT is funded by NSERC (grant no.s RGPIN-2016-05277 and RGPIN-2022-03589), and the Institute for Biodiversity, Resilience, and Ecosystem Services (UBC Okanagan).  RCT would also like to acknowledge the contributions of undergraduate researcher Kim Wilcott who contributed to our understanding of earlier versions of the model.  
The research of SW was partially supported by the EvoGamesPlus Innovative Training Network funded by the European Union’s Horizon 2020 research and innovation programme under the Marie Skłodowska-Curie grant agreement No 955708.
RCT and SW also acknowledge the contribution provided by BIRS through workshop 19W5108 where they met.

\printbibliography
\appendix{}
\section*{Definitions of key terms}

\begin{itemize}
    \item {\it Phase-tipping (P-tipping):}
    { In cyclic systems, phase-tipping is a critical transition, or a regime shift, from a cyclic base state to an alternative stable state, due to a (sudden) change in the external input, that occurs only from certain phases of the cycle.
    }
    \item {\it Climate variability:}
    Climate variability is the variation of climate conditions, such as temperature and precipitation, measured as deviations from the mean, on timescales
    of 1 to 10 years.
    \item {\it Climate reddening}
     As the time series of climate metrics become more autocorrelated, the distribution of frequencies in these time series shifts so that the lower frequencies become more dominant.  This downward shift in the frequency spectrum is called "climate reddening".
    \item {\it Basin of attraction of the limit cycle:}
    The set of all initial states 
    that converge to the limit cycle over time, for a fixed value of the system parameters.
    \item {\it Escape event:} 
    We say the  oscillatory system has escaped the limit-cycle base attractor if, following a sudden change 
    { to a new value of $r(t)$}, the state of the system lies outside the basin of attraction of the limit cycle {for this new value of $r(t)$}.
    \item {\it Rescue event:} 
    Once outside the basin of attraction of the limit cycle, the system moves toward the extinction state. However, within a short window of opportunity,  the system can be rescued by another change in $r(t)$ that positions the system state back inside the basin of attraction of the limit cycle. 
    \item {\it Phase of the cycle:}
    For every point in the basin of attraction of a limit cycle, one can define a phase of the oscillation.  For the simple cycles we study here, one can think of the phase as the angle measured from some reference axis, see \citep{alkhayuon:2021}. More generally, one could use the concept of isochrones \citep{guckenheimer1975isochrons}.
 \end{itemize}

\begin{sidewaystable}
{
    \caption{Parameter values for the two predator-prey models applied to the Canada lynx and snowshoe hare predator-prey system.  All of the parameter values, except the Allee parameters, are taken from  \citet{strohm:2009,tyson:2009}.}
    \label{tbl:params}
    \begin{center}
        \begin{tabular}{ c c c c c}
            Parameter  & Units         & RMA model & MayA model\\
            \hline
            $r$        &  1/yr                 & $[0,3]$   & $[0,4]$ & {prey productivity rate} \\
            $c$        &  ha/(prey$\cdot$yr)   & $0.19$  & $0.22$ & density-dependent competition parameter\\
            $\alpha$   &  prey/(pred$\cdot$yr) & $800$     & $505$ &  predator saturation kill rate  \\
            $\beta$    &  prey/ha              & $1.5$     & $0.3$  &  predator kill half-saturation constant.\\ 
            $\chi$     &  pred/prey            & $0.004$   & n/a    &  prey-to-predator conversion ratio \\
            $\delta$   &  1/yr                 & $2.2$     & n/a    &  predator
            mortality rate\\
            $s$        &  1/yr                & n/a       & $0.85$  & low-density
            predator growth rate\\
            $q$        &  prey/pred            & n/a       & $205$  &   the minimum prey-to-predator biomass ratio \\
            $\mu$      &  prey/ha              & $0.03$    & $0.03$ &  The Allee threshold parameters \\  
            $\nu$      &  prey/ha              & $0.003$   & $0.003$ & The Allee saturation parameters \\
            $\epsilon$ &  prey/ha              & n/a       & $0.031$ &  predator minimum density parameter \\
            \hline
\end{tabular}
\end{center}
}
\end{sidewaystable}
\end{document}